\documentclass{emulateapj}
\usepackage{natbib}
\usepackage{amsmath}
\usepackage{ulsy} 
\usepackage{subfigure}
\usepackage{soul,color}

\newcommand{\pfrac}[2]{\left(\frac{#1}{#2}\right)}
\newcommand{\icarus}{Icarus}
\newcommand{\pasa}{PASA}

\shorttitle{Constraining the Protosolar Nebula}
\shortauthors{Kretke et al.}

\begin{document}
\title{A Method to Constrain the Size of the Protosolar Nebula}
\author{K. A. Kretke, H. F. Levison, and M. W. Buie}
\affil{Southwest Research Institute, Department of Space Studies, 1050 Walnut Street, Suite 400, Boulder, CO 80302, USA}
\email{kretke@boulder.swri.edu}
\and
\author{A. Morbidelli}
\affil{Observatoire de la C\^ote d'Azur, BP 4229, 06304 Nice Cedex 4, France}

\begin{abstract}
Observations indicate that the gaseous circumstellar disks around young stars vary significantly in size, ranging from tens to thousands of AU. Models of planet formation depend critically upon the properties of these primordial disks, yet in general it is impossible to connect an existing planetary system with an observed disk. We present a method by which we can constrain the size of our own protosolar nebula using the properties of the small body reservoirs in the solar system. In standard planet formation theory, after Jupiter and Saturn formed they scattered a significant number of remnant planetesimals into highly eccentric orbits. In this paper, we show that if there had been a massive, extended protoplanetary disk at that time, then the disk would have excited Kozai oscillations in some of the scattered objects, driving them into high-inclination ($i \gtrsim 50^\circ$), low-eccentricity orbits ($q \gtrsim 30$ AU). The dissipation of the gaseous disk would strand a subset of objects in these high-inclination orbits; orbits that are stable on Gyr time scales. 
To date, surveys have not detected any Kuiper Belt Objects with orbits consistent with this dynamical mechanism.
Using these non-detections by the Deep Ecliptic Survey (DES) and the Palomar Distant Solar System Survey we are able to rule out an extended gaseous protoplanetary disk ($R_D\gtrsim 80$ AU) in our solar system at the time of Jupiter's formation.
Future deep all sky surveys such as the Large Synoptic Survey Telescope (LSST) will all us to further constrain the size of the protoplanetary disk.
\end{abstract}

\section{Introduction}
Understanding the properties of the gas rich disks around young stars is crucial to understanding the formation of planetary systems.
Observations demonstrate that these protoplanetary disks exhibit significant diversity in their sizes, masses, and lifetimes \citep[e.g.][]{Haisch.etal.2001,Mamajek.2009,Andrews.etal.2009,Andrews.etal.2010}.
These variations must affect the ubiquity and diversity of observed exoplanet systems.
However, it impossible to connect any of the Gyr-old existing planetary systems directly to the observed few Myr-old disks.
However, our own solar system provides a unique test case.  
In this one system we have access to a significant amount of data, stored in the small body reservoirs, that allows us to make inferences about the early state of our own solar nebula.

There is a long tradition of using observations from the present day solar system to try to constrain the properties of the original solar nebula.
Early on, researchers noted that they could define a minimum mass solar nebula (MMSN), the lowest mass disk that could have created the planets in their current locations, by augmenting the observed solids in the planets with hydrogen and helium until they reach solar composition \citep{Weidenschilling.1977a,Hayashi.1981}.
The resulting power-law profile has served the standard benchmark model in planet formation simulations.
More recently, researchers have suggested that an apparent hard edge to the classical Kuiper belt at 50 AU \citep{Allen.etal.2001} may be indicative of the initial size of the protoplanetary disk. 
Indeed, \citet{Adams.2010} and references within have used this apparent outer edge to the solar system as part of an effort to understand the birth environment of our solar system.
For example, \citeauthor{Adams.2010} presents arguments that a star cluster dense enough to truncate the disk at around 50 AU is consistent with a cluster capable of producing a super-nova able to contaminate our early solar system with short lived radioactive isotopes.
Although this paints a self-consistent picture, it is important to note that the mechanisms that truncate the outer regions of the disk (especially photoevaporation) actually act primarily on the disk \emph{gas}, while the data (the distribution of planets and KBOs) refers to the disk \emph{solids}.
There are well-known physical mechanisms capable of decoupling the solids and the gas \citep{Weidenschilling.1977,Youdin.Shu.2002}.
Indeed, some authors have even argued that a large reservoir of drifting solids born in an extended gaseous disk could be useful in enhancing the solid to gas ratio, increasing the likelihood of planetesimal formation and core formation \citep[e.g.][]{Youdin.Chiang.2004}.
Therefore one cannot claim that an ``edge'' in the planetesimal disk necessarily corresponds to an edge to the gas disk.

In this paper we present an argument that constrains the size of the gaseous disk at the time of Jupiter's formation.
Planet formation is an inefficient process; current models suggest that the total mass of planetesimals left over after the planets are fully assembled is equivalent to, or up to 5 times greater than, the total mass of solids incorporated into the planets (e.g. \citealt{Pollack.etal.1996,Lissauer.1987,Thommes.etal.2003,Alibert.etal.2005,Hubickyj.etal.2005} and see \citealt{Thommes.Duncan.2006} for a review).
Therefore during and shortly after the formation of the giant planets, 
must have been a significant number of planetesimals left over in the planets' feeding zones.
These planetesimals are fated to undergo close encounters with the planets, and many will be scattered to high eccentricity orbits.
In a simple 3-body scattering, the perihelion of the scattered object remains roughly unchanged, meaning that the planetesimal will return to undergo additional close encounters.
Eventually these objects will either collide with the planet or be ejected by the system.

However, in order to form the planets there must also have been a circumstellar gas disk, a disk that potentially could have extended beyond the orbits of the planets.
And the presence of a massive gaseous disk can alter these scattered planetesimals' orbits.
For planetesimals between a few tens to a few thousands of kilometers in size, they are too large to experience significant gas drag\citep{Adachi.etal.1976}, and yet are too small to undergo any sort of tidal interaction with the disk \citep[][and references within]{Ward.1997}.
Given the cylindrical symmetry of the disk, the major gravitational effect of the disk is of the Kozai type \citep{Kozai.1962}, which arises when a perturber can  be treated as a circular ring.
In the presence of the potential from a massive gaseous disk, planetesimals scattered from the Jupiter/Saturn region into moderate inclination, high eccentricity orbits may undergo Kozai oscillations.

During a Kozai oscillation, the presence of perturbing material exterior to the scattered particle forces the particle exchange eccentricity for inclination, all the while conserving the z-component of the angular momentum.
In the situation presented here the gaseous disk serves as the perturbing potential.
The scattered object oscillates between a low-inclination, eccentric orbit and an high-inclination, more circular orbit. 
During the high eccentricity phase the particle is susceptible to additional close encounters with the planets, but during the low eccentricity phase, the particle's orbit will no longer cross the planets.
After the disk disperses, statistically some of the objects will be stranded at high inclination orbits. 
This dynamical mechanism is capable of producing objects with large enough perihelia to place them in Kuiper Belt on orbits that are stable for the lifetime of the solar system with the planets in their present day orbital configuration.
But while these particles have semi-major axes that would place them in the Kuiper Belt, they would have substantially higher inclinations than what has been observed so far.
In this paper we explore this dynamical mechanism as a way to use the current distribution of Kuiper Belt Objects to place constraints on the early structure of the protoplanetary disk.

In \S\ref{sec:model} we describe our numerical method and model parameters.
In \S\ref{sec:results} we discuss the outcomes for various disk parameters.
In \S\ref{sec:observations} we discuss how we can use these results to constrain the size of the protosolar nebula using results from the Deep Ecliptic Survey \citep[DES;][]{Millis.etal.2002,Elliot.etal.2005,Gulbis.etal.2010} and the Palomar Distant Solar System Survey \citep{Schwamb.etal.2009,Schwamb.etal.2010}.
And finally, in \S\ref{sec:conclusions} we summarize our results and discuss the implications.

\section{Model}\label{sec:model}
In this paper we investigate the behavior of planetesimals initially in the giant planet regions and model their gravitational interactions with the planets and the gaseous disk.
We initialize these simulations with four giant planets in the compact configuration proposed by \citet{Tsiganis.etal.2005} as the initial positions of the planets in the \emph{Nice Model}, with Jupiter, Saturn, Uranus and Neptune at 5.45, 8.18, 11.5, and 14.2 AU, respectively.
However, as we discuss in \S \ref{sec:results}, the results of this paper are insensitive to this choice.
For each simulation we calculate the orbital evolution of $N_{\rm TP} = 6\times 10^3$-$10^4$ massless test particles initially distributed from 4-15 AU with small eccentricities ($e=0.1$) and inclinations ($0^\circ<i<3^\circ$) to represent the planetesimals left over in the planet forming region.
We calculate the orbital evolution of the scattered planetesimals using the SWIFT RMVS3 numerical integrator package \citep{Levison.Duncan.1994}.
We integrate the orbits for 5 Myr with a time-step of 0.15 yr, 
and we remove particles from the simulation if they become unbound or they experiences a collision with a planet or the Sun.
Additionally, as we are interested only in particles whose orbits are not perturbed by the potential due to the birth cluster or the galaxy, we remove particles if their semi-major axis ($a$) becomes larger than 300 AU.

To calculate the particles' gravitational interaction with the disk gas we numerically integrate Poisson's equation over a cylindrically symmetric grid.
This allows us to obtain the potential and the accelerations derived from the disk gas on the particles using bicubic interpolation. 
For computational simplicity we use simple a simple parametric form for the disk gas. 
We describe the surface density profile as a function of distance from the sun $r$ at time $t$ as a truncated power-law, 
\begin{equation}
	\Sigma(r,t) = \Sigma_0 \pfrac{r}{\rm AU}^{-\gamma} \Theta(r,R_D) \exp\pfrac{-t}{\tau_D},
	\label{eq:Sigma}
\end{equation}
where $\Theta(r,R_D)$ describes the functional form of the truncation of the outer disk at size $R_D$, and $\tau_D$ is the disk depletion timescale. 
For the fiducial disk models we use $\Sigma_0=2000 {\rm g~cm}^{-2}$ and $\gamma=3/2$ consistent with the benchmark minimum mass nebula profile \citep{Hayashi.1981}.
We investigate the effect of the sharpness of the disk truncation on distribution of planetesimals by comparing two truncation profiles, a sharp cutoff,
\begin{equation}
	\Theta(r,R_D) = \Theta_{\rm cut}(r,R_D) \equiv \left\{ \begin{array}{rl}
 1, &\mbox{ if 1 AU $\le r \leq R_D$,} \\
 0, &\mbox{ otherwise}.
       \end{array} \right.
\end{equation}
and an exponential cutoff 
\begin{equation}
\Theta_{\rm exp}(r,R_D) \equiv \left\{ \begin{array}{rl}
 \exp(-r/R_D), &\mbox{ if $r \ge 1$ AU,} \\
 0, &\mbox{ otherwise}.
 \end{array} \right.
\end{equation}
\citep[c.f.][]{Andrews.etal.2009,Andrews.etal.2010}.
In all disk models we truncate the inner disk at 1 AU, but we vary the outer disk cutoff ($R_D$) from 30 AU to 100 AU.
To preserve the stability of the integration we use a ``puffy'' exponential profile for the disk vertical profile,
\begin{equation}
	\rho(r,z) = \Sigma(r) \exp\left(-\frac{z}{H}\right),
\end{equation}
where $H$ is constant throughout the disk.
This disk model is chosen for its simplicity and is not intended to represent a detailed disk evolution model.

There are five free parameters for the disks in these simulations: $\Sigma_0$, $\tau_D$, $R_D$, and $\gamma$, and the functional form of $\Theta(r,R_D)$.
In total we present the results of the evolution in 35 different gas disks.
Our fiducial parameters are $\Sigma_0=2000~{\rm g~cm^{-2}}$, $\tau_D = 2$ Myr, $\Theta(r,R_D)=\Theta_{\rm cut}$, $R_D=100$ AU, and $\gamma=3/2$.
With the fiducial disk parameters, we perform 7 simulations varying $\Sigma_0$ from 20 to 4000~${\rm g~cm^{-2}}$.
We also perform 6 simulations varying $\tau_D$ from $2\times10^4$ to $4\times10^6$ years. 
Using the fiducial $\Sigma_0$ and $\tau_D$ we then perform simulations with $R_D=$30, 40, 50, 60, 80 and 100 AU.  
For the sharp cut off we allow $\gamma = 3/2$ (the MMSN value) and $\gamma=1$ (a value consistent with some viscous evolution models).
For the exponential cutoff we use $\gamma = 1/2$ and $1$, appropriate values for observed disks \citep{Andrews.etal.2010}.
In addition, we compute one simulation in the fiducial disk with the four giant planets in the configuration of the modern day solar system.
We run these scattering simulations for 5 Myr (10 Myr for the disk with $\tau_D=4\times 10^6$).

In most of these simulations, some particles are captured by the Kozai
resonance and forced into orbits with $q>30$ AU; orbits that are
potentially stable in the modern day solar system. 
However, although a large number of particles undergo oscillations with large perihelion
values at some point during their orbital evolution, at the end of the
scattering simulations only a handful of these particles happen to have been 
caught on the $q$ phase of their orbits. 
Still, the time at which a given particle is scattered into
one of these orbit is random, and could as easily have happened right before the disk dispersed as at any other time. 
Therefore instead of restricting our stability calculation to only those particles still active at the end of the simulations, we randomly extract the orbital elements ($a$,$e$, and $i$) of particles with large perihelia ($q>20$ AU) values throughout the simulations.
We then randomly assign orbital angles and create a new suite of equally probable, potentially stable particles.
We integrate these new particles in the presence of the four giant planets, in their modern day configuration, for 4.5 Gyr. We use these particles to
define the probability that a particle will survive as a function of
its perihelion distance at the beginning of the stability calculations
($P_S(q)$). 
We can then return to the initial scattering calculations
and use the probability that each particle (at its given perihelion
distance) will be stable to calculate the expected fractional
retention of particles,
\begin{equation}
f \equiv \sum_{i=1}^{N_{\rm TP}} P_S(q_i)/N_{\rm TP}.
\label{eq:P_S}
\end{equation}
Here $P_S(q)$ is the function defined by the stability calculation,
and $q_i$ refer to the perihelion distance of each particle at the end of the initial scattering simulations. 
Because $P_S(q)$ goes to zero as $q_i$
approaches 20 AU in all simulations, we are confident that the initial
cut-off of $q=20$ AU does not impact the statistics in any part of
this study. To determine our 95\% confidence intervals, we assume that
our data is distributed according to a binomial distribution and use a
non-informative, Bayes-Laplace uniform prior \citep[see][for a nice
description of this method]{Cameron.2011}.

In these simulations, the particles only interact with the gas though its gravitational potential, not through aerodynamic drag.
As indicated in \citet{Brasser.etal.2007} the gas drag is only dynamically important for objects with diameters smaller than about $D =30$ km.  
The impact of the gas drag on these small bodies is to circularize the orbits, and thus they do not participate in the effect explored in this paper.
Furthermore, as we discuss in \S \ref{sec:observations}, most of the mass is likely to be in 100 km, or larger, sized objects and smaller objects would not be detectable in current datasets.
Therefore we only concern ourselves with these larger bodies and neglect gas drag in these simulations.


\section{Results}\label{sec:results}
\begin{figure}
\includegraphics[width=\columnwidth]{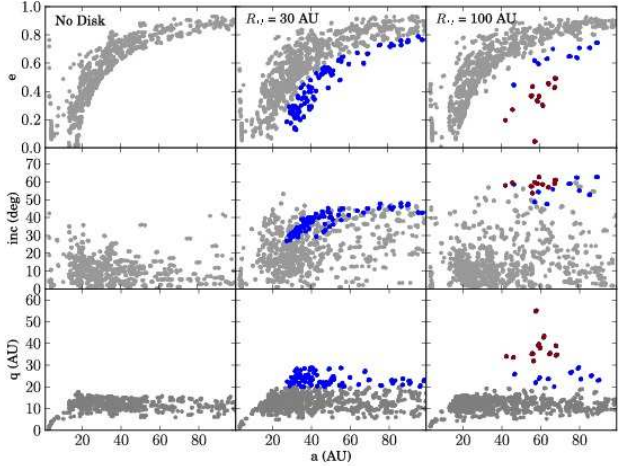}
\caption
{The eccentricity (top), inclinations (middle), and perihelion distance (bottom) for particles after 5 Myr in simulations with no disk potential (left), and the potential of a MMSN disk truncated at 30 AU disk (middle) and 100 AU (right).
In blue we show particles stable in the compact early solar system and in dark red we show particles stable in the present day solar system, while test particles on unstable orbits are in gray.
}
\label{fig:init}
\end{figure}

In figure \ref{fig:init} we compare the orbital distribution of planetesimals produced by disks of various sizes after the scattering simulations, by which time the gas disks have dispersed.
Without the gravitational potential of a gaseous disk (left-hand panels), the planetesimals are scattered by the planets into eccentric and moderately inclined orbits, but the perihelia ($q$) remain roughly unchanged.
Therefore, all of the particles remain on planet-crossing orbits that are unstable on timescales short compared to the age of the solar system.
These unstable particles are indicated by gray points.
In contrast, in middle and right panels we show the particles' evolution in the presence of a disk with a MMSN profile ($\gamma=3/2$) and a sharp cutoff ($\Theta=\Theta_{\rm cut}$), truncated at $R_D=$ 30 AU and 100 AU, respectively.
In both of these simulations, a subset of the particles are placed into orbits with larger perihelia and higher inclinations.
In the small disk, some particles have their perihelia raised so that they have a high probability of being stable in the early compact solar system, with $q>20$ AU (particles marked in blue), but it is only in the larger disk that the perihelia are raised to the point where they would be stable in the modern solar system, $q>35$AU (marked in red).

\begin{figure*}
\includegraphics[width=\textwidth]{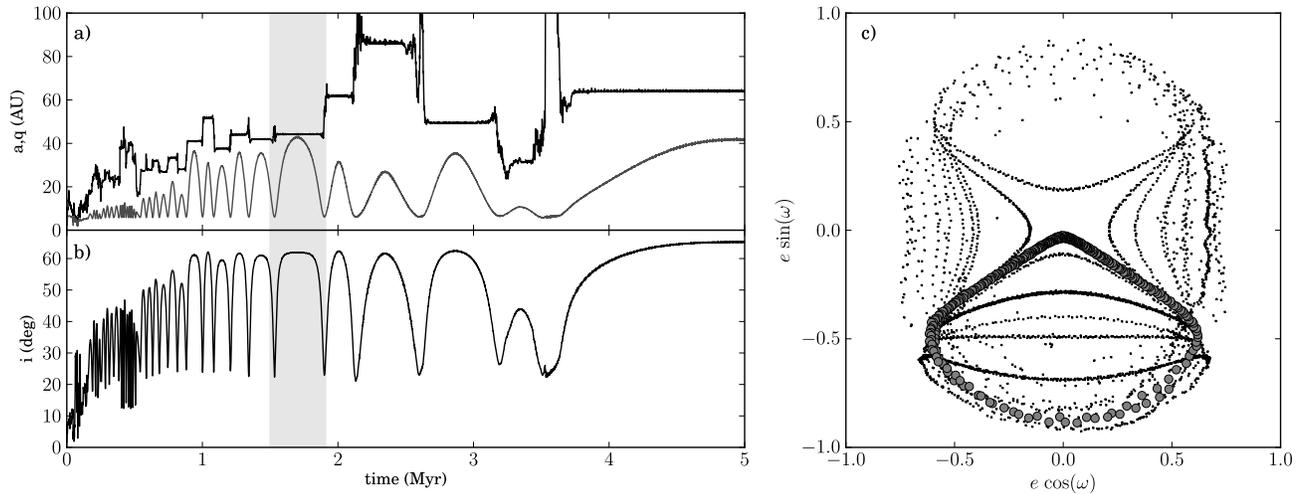}
\caption{
In panel(a) we show the semi-major axis (black curve) and pericenter distance (Grey curve) for an example particle in the fiducial simulation.  
This particle is scattered by Jupiter and then undergoes a series of Kozai oscillations and close encounters.  
Eventually it is stranded in a high inclination orbit.
In panel (b) we show the inclination of the particle.
The shaded region indicates an example Kozai cycle.  
In panel (c) we show the eccentricity and argument of pericenter in a polar plot for the example particle.  
The small dots show these parameters at all times, while the large circles indicate quantities while the particle is in the shaded region of the left panels.
}
\label{fig:kozai_end}
\end{figure*}

These particles get onto these high inclination, low eccentricity orbits via Kozai oscillations.
In figure \ref{fig:kozai_end} we show the temporal evolution of an example particle in the 100 AU disk.
The particle undergoes a series of close encounters with Jupiter and is scattered out to a high eccentricity orbit.
After about half a million years we see a number of intervals in which the perihelion distance increases and decreases in phase with oscillation in the inclination while the semi-major axis remains constant, as is characteristic for Kozai oscillations.
As further evidence, in the right hand panel we plot the eccentricity and the argument of perihelion for this particle.
The large circles indicate a time interval (shaded in the left hand panel) of a single Kozai cycle.  
In this time the argument of perihelion oscillates, indicating that this is a true Kozai cycle.
Additionally, by comparing figure  \ref{fig:kozai_end}c with the Kozai dynamics exerted by the planets on objects with larger semi-major axes \citep[c.f.][]{Thomas.Morbidelli.1996} we can be confident that the perturbations seen here are dominated by the \emph{external} disk potential not perturbations from the interior planets.
During each of the periods of high eccentricity (small $q$) the particle undergoes additional encounters with Jupiter and Saturn, causing sharp jumps in the particle's semi-major axis.
As the disk is depleted the timescale for the oscillation become longer. 
In the end the period of oscillation becomes infinite, stranding the particle on a high inclination orbit.

\begin{figure*}
	\subfigure[sharp cutoff and $\gamma=3/2$]{\includegraphics[width=0.5\textwidth]{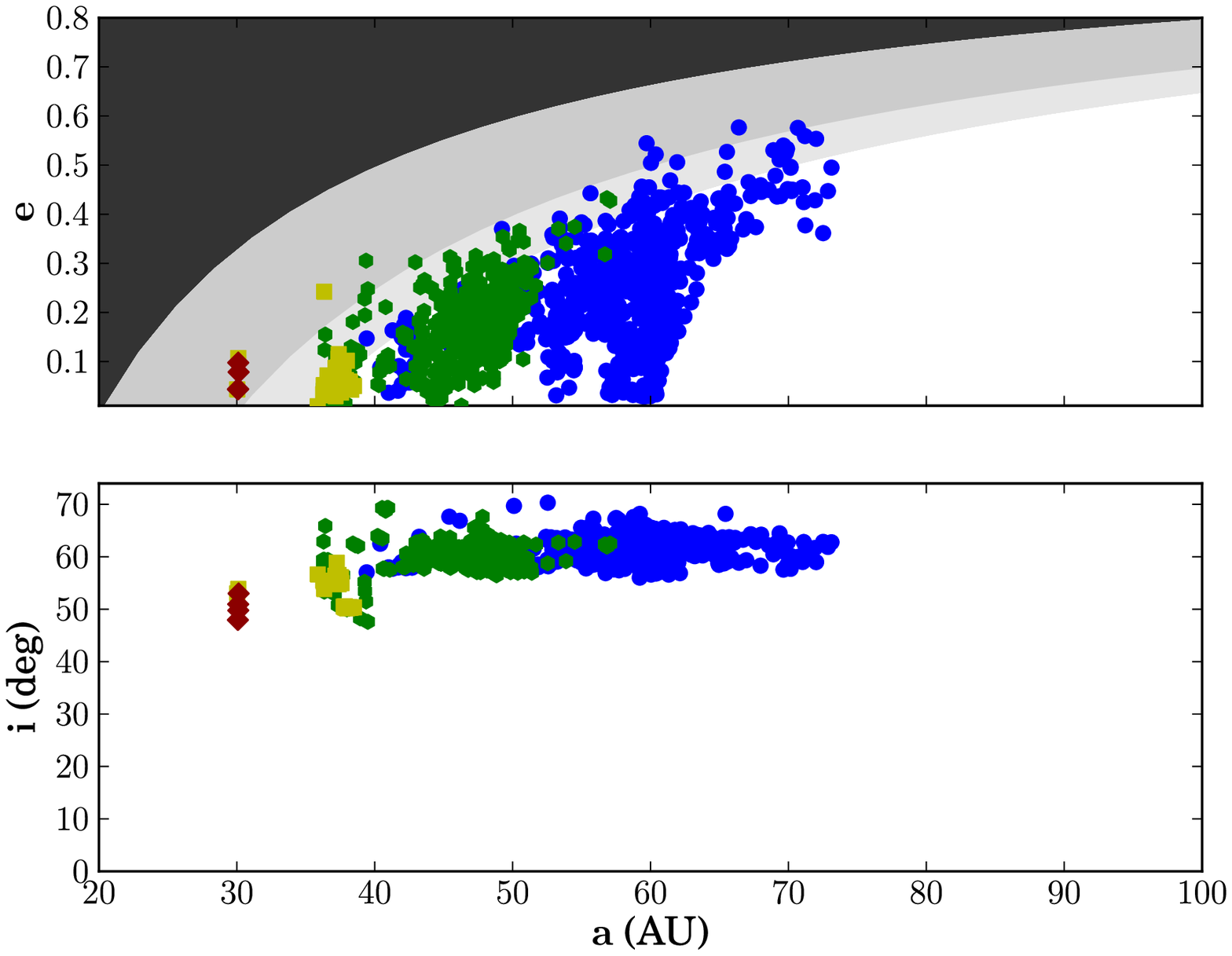}\label{fig:StabFig1}}
	\subfigure[sharp cutoff and $\gamma=1$.]{\includegraphics[width=0.5\textwidth]{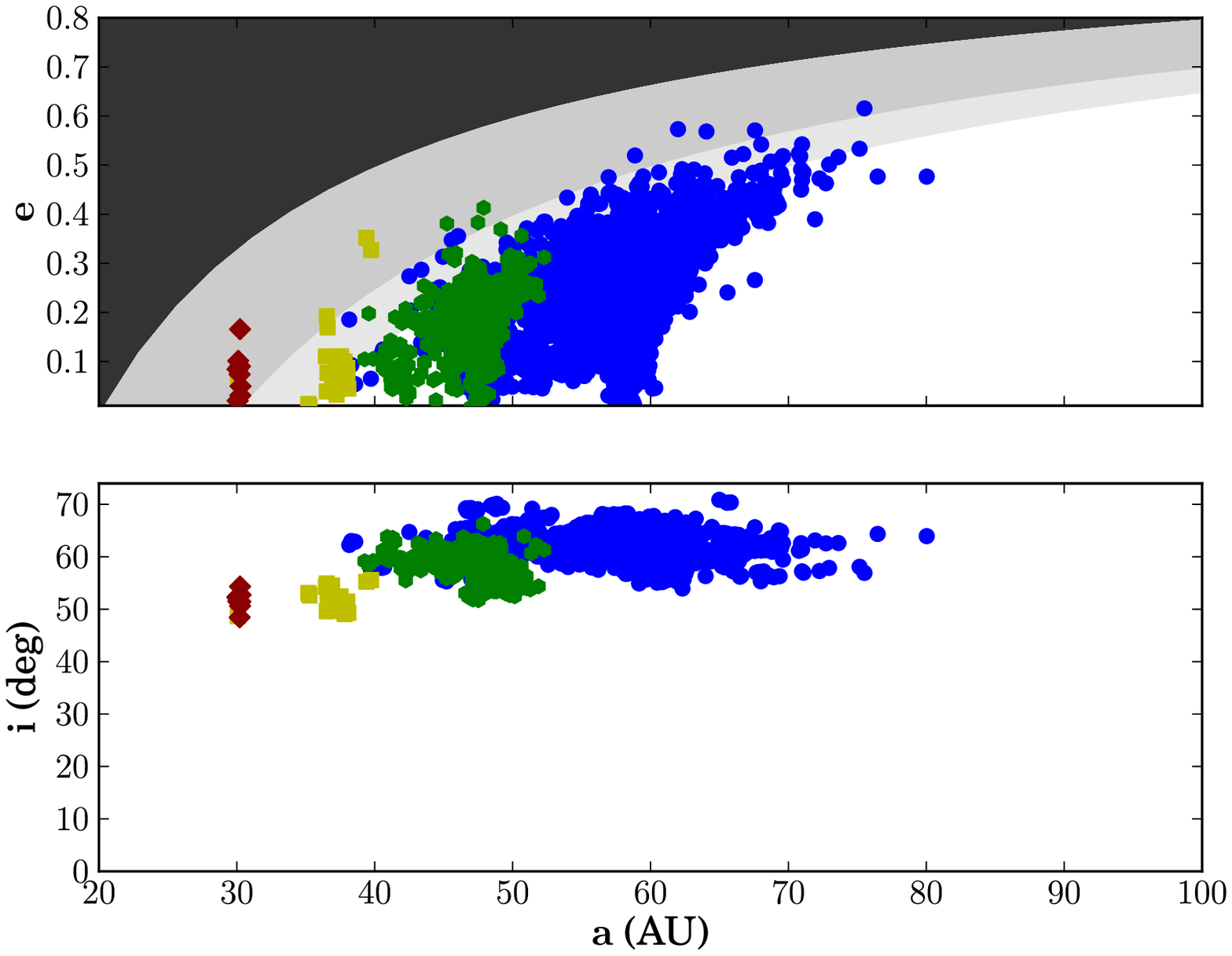} \label{fig:StabFig2}}
	\subfigure[exponential cutoff and $\gamma=1/2$.]{\includegraphics[width=0.5\textwidth]{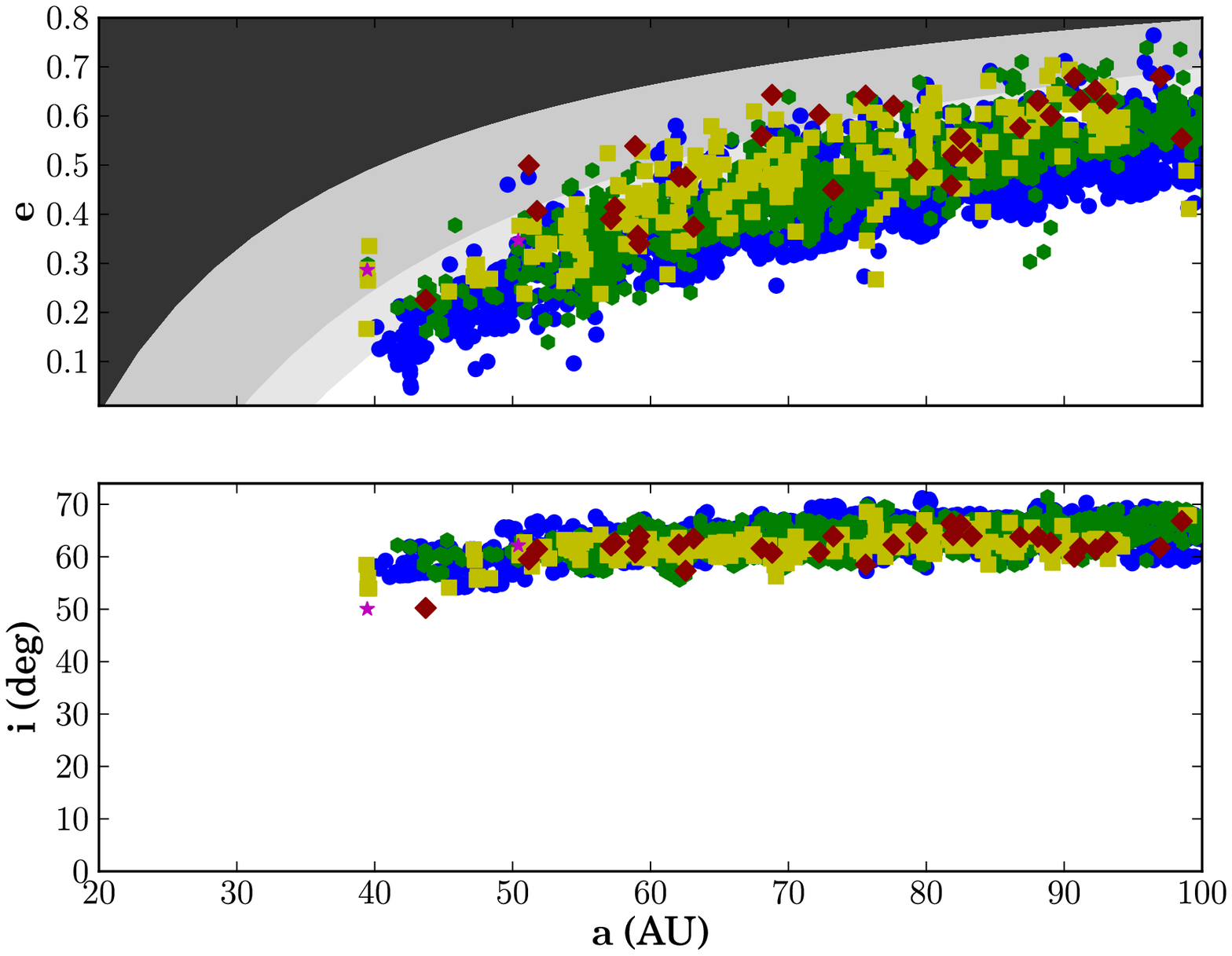} \label{fig:StabFig4}}
	\subfigure[exponential cutoff and $\gamma=1$.]{\includegraphics[width=0.5\textwidth]{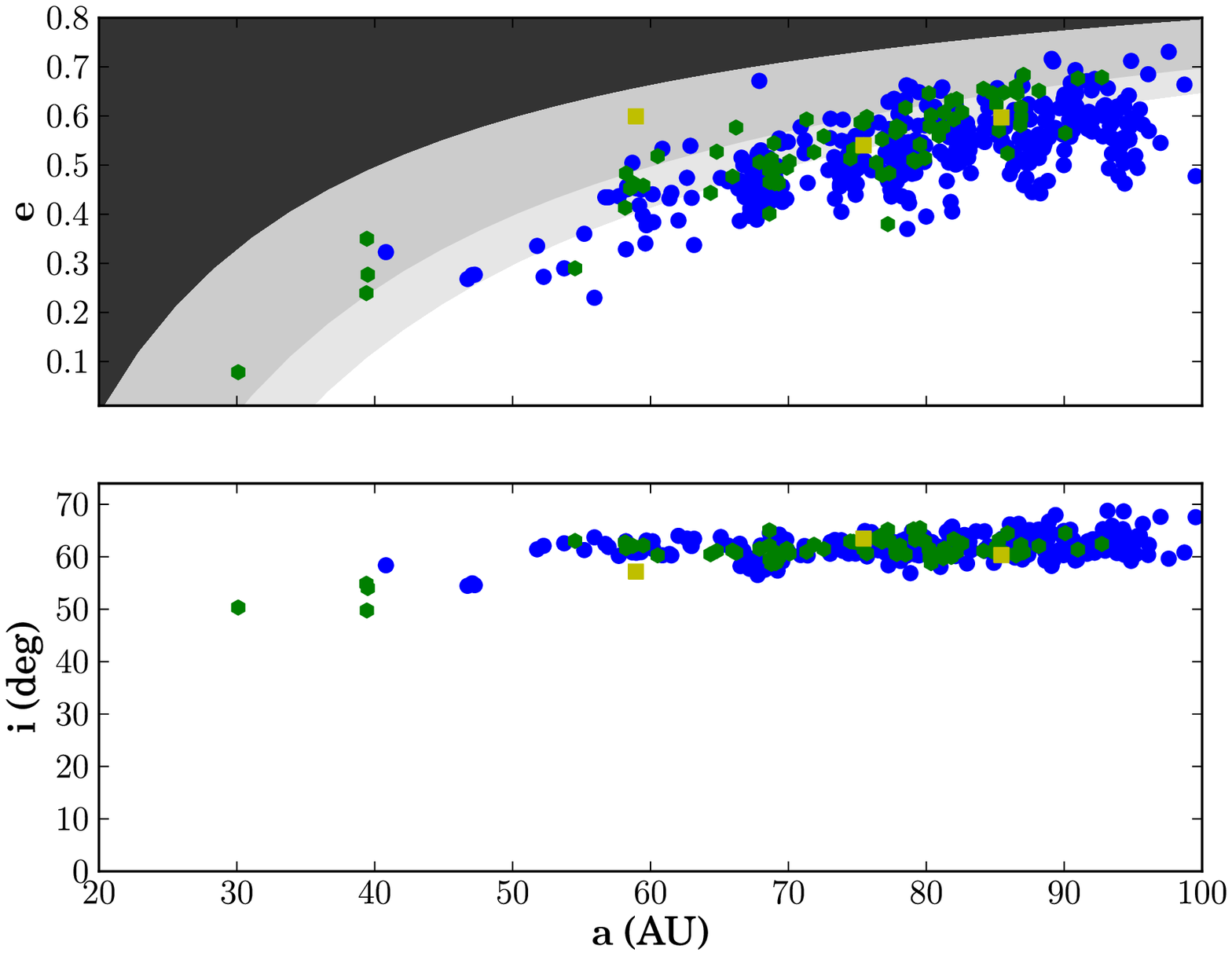} \label{fig:StabFig3}}
\caption
{The eccentricity and inclination of objects left in disk with disk size $R_D=$ 40, 50, 60, 80, and 100 AU (magenta stars, red diamonds, yellow squares, green hexagons, and blue circles, respectively). In the upper panels we shade the regions according to perihelia distance ($q=20,30,35$ AU darker to lighter).}
\label{fig:StabFig}
\end{figure*}

Regardless of the exact initial orbits of the planets, at some point they must have moved into their currently observed configuration.
Therefore we are really only concerned with the orbital distribution of particles that would be observable today due the fact that their orbits are stable in the present day solar system, for the lifetime of the solar system.
The location of all stable objects produced in simulations with various disk profiles and sizes are shown in figure \ref{fig:StabFig}.
We show four panels, a ``shallow'' and ``steep'' surface density profile for each of our two truncation functions.
For the upper plots (sharp cutoff) the disk truncation function is a step function ($\Theta = \Theta_{\rm cut}$).
The surface density profiles are, for panel a, the benchmark MMSN $\gamma=3/2$ and, for panel b) the shallower $\gamma=1$ more consistent with viscous evolution.
For the lower plots (exponential cutoff) the disk truncation function is $\Theta=\Theta_{\rm exp}$.
We choose to show shallower surface density indices, $\gamma=1/2$ and $\gamma = 1$ in panels c and d, respectively, to be consistent with observations of disks \citep{Andrews.etal.2010}.
The points are color-coded by the disk size, with $R_D=$ 40, 50, 60, 80, and 100 AU in magenta stars, red diamonds, yellow squares, green hexagons, and blue circles, respectively.

The orbital distribution of stable test particles are qualitatively different depending on the size and cutoff profile of the disk.
In disks with sharp cutoff there is a limited semi-major axis range in which stable high-inclination particles can be produced.
An exterior quadrapole moment is much more effective in raising the Kozai resonance than an inner one, and the dynamics of the two are significantly different \citep{Thomas.Morbidelli.1996}.
Therefore larger disks can excite Kozai oscillations in particles with larger semi-major axes.
Additionally, these disks are well approximated by an annulus at the outer edge as that is where the mass is concentrated.
The shallower surface density profile ($\gamma=1$) is slightly better at producing high inclination particles as more of the disk mass is at this outer radii.
In disks with exponential profiles there is mass at a large range of radii.
This allows particles to undergo Kozai oscillations at large $a$,
however the fact that the disk is not dominated by a single annulus of material at the outer edge means that the oscillations are shallower.
Therefore the objects produced do not have very low eccentricity orbits. 
The shallower surface density profile ($\gamma=1/2)$ again more effectively generates these particles due to the fact that there is more mass at larger radii to excite Kozai oscillations.

\begin{figure}
\includegraphics[width=\columnwidth]{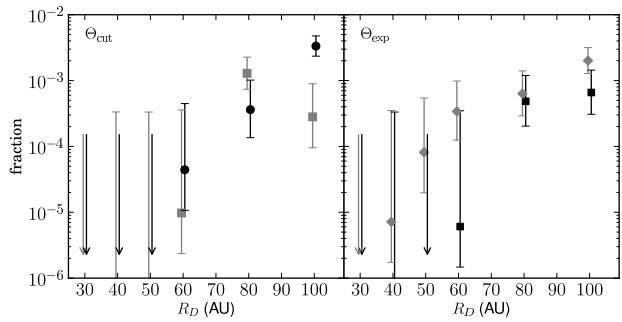}
\caption
{The fraction of objects left in stable high-inclination orbits after 5 Myr as a function disk size and profile for a disk with a sharp cutoff (left), and an exponential cutoff (right). The circles, squares, and diamonds correspond to $\gamma = 3/2, 1,$ and $1/2$, respectively.  The downwards arrows indicate simulations in which no particle ever has its perihelia raised to a potentially stable orbit.
}
\label{fig:frequency_shape}
\end{figure}

The effect of disk size on retention can be seen more clearly in figure \ref{fig:frequency_shape} in which we show the fractional retention ($f$, defined in eq.~\ref{eq:P_S}) as a function of the disk size and profile (using the parametric form shown in equation (\ref{eq:Sigma})).
The downwards arrows indicate the upper limit on simulations in which no particle ever has its perihelia raised to a potentially stable orbit.
There is a trend that larger disks tend to retain a larger fraction of particles.

\begin{figure}
\includegraphics[width=\columnwidth]{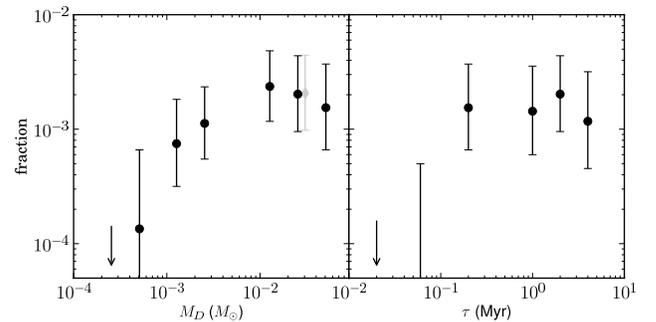}
\caption
{The fraction of objects left in stable high-inclination orbits as a function disk mass (left), and disk lifetime (right) for a disk with $R_D=100 AU$ and $\gamma = 3/2$.
In the left-hand plot the light gray diamond indicates the results for a simulation in which the scattering is done by planets on their present day orbits.} 
\label{fig:frequency}
\end{figure}

We note that in these various simulations we maintained the same normalization of the surface density profile of the gas disk at 5 AU ($\Sigma_5= 180 {\rm g~cm}^{-2}$).
However, this arbitrary choice does not significantly impact the results.
In figure \ref{fig:frequency} we show the fraction of particles retained as a function of disk mass (with the fiducial depletion timescale of 2 Myr) and depletion timescale (with the fiducial disk mass of 0.05 $M_\odot$).
Other than at very small disk masses and very short disk lifetimes the fraction of retained particles in remarkable insensitive to these disk parameters.
In order to understand this insensitivity, we examine the timescale for a Kozai oscillation.
For $a<<R$, the characteristic timescale for Kozai oscillations due to a perturbation of a narrow, infinitesimally thin annulus of radius $R$ and mass $M$ on a particle or semi-major axis $a$ is
\begin{equation}
	\tau_K \sim \frac{2}{3\pi} \frac{M_\sun}{M} \pfrac{R}{a}^{3/2} \sqrt\frac{R^3}{G M_\sun}
	\label{eq:tauK}
\end{equation}
(e.g. \citet{Kiseleva.etal.1998}).
Despite the fact that the systems considered in this paper are more aptly described as a series of concentric rings with finite vertical thickness, the Kozai timescale for the disks with a $\Theta=\Theta_{\rm cut}$ can be reasonably well approximated by modeling the disk as an thin annulus of mass $M=M_D$ and radius $R=R_D$.
This timescale becomes longer as $R_D$ increases, and as $M_D$ decreases.
Therefore there is a limiting mass for which a single oscillation can occur within the disk lifetime.
For disks with a timescale for dispersal of less than this we do not see any particles excited into orbits of interest.
As mentioned previously, for our fiducial disk of a MMSN truncated at 100 AU, this estimate of the Kozai timescale yields $2\times 10^4$ years.  
Additionally, for the fiducial disk lifetime of 2 Myr, the minimum mass required in order to allow for a single Kozai oscillation is $2\times 10^{-4}$ $M_\odot$.
This means that for our fiducial disk parameters, the particles undergo \emph{many} Kozai cycles, and during each close encounter with a planet the particle may either be scattered into or out of the Kozai cycle.
As the disk mass or lifetime increases, a larger fraction of the particles undergo Kozai oscillations at some point in their evolution.
However, most of these particles at some point undergo a close encounter with a giant planet and are ejected from the system.
Therefore the number of particles left after disk dispersal asymptotes with increasing disk mass.
So long as Jupiter and Saturn did not form so late in the disk evolution that, within a few tens of thousands of years, the disk mass decreased to less than a Jupiter Mass, we are insensitive to the actual disk mass and lifetime.
This means that we can safely assume we are in a parameter space in which disk size and profile are important parameters, but disk mass and lifetime are not.

One may also be concerned about the sensitivity of these results on our assumption about the position of the planets.
There are two possible reasons the planetary positions may be important, first the secular perturbations from the interior planets may stabilize the particles against the Kozai resonance.  
Additionally, the location of the planets may scatter the particles in different ways.
Therefore, we perform the same calculation with the planets in the present-day solar system, as opposed to the {\it Nice} initial conditions.
The fractional trapping is virtually identical in this case (see the diamond in fig \ref{fig:frequency}).
We note that in the fiducial disk model, the total mass of the disk gas interior to 30 AU is over 13 $M_{\rm Jup}$, therefore the secular influence of the planets is minor compared to that of the disk.
This is because, in order to undergo a deep Kozai oscillation, a particle at a given semi-major axis must either have a minimum eccentricity or inclination.
As we are interested in using the Kozai resonance to produce high perihelion (low eccentricity) objects, we are interested in the initially eccentric objects.
In order for the initially dynamically cold particles need to reach these minimum eccentricities, they must undergo a close-encounter with a planet. 
After the scattering the particles orbit will have a perihelia close to the orbit of the scattering planet ($q = a (1-e)\sim a_p$).
Therefore, if we look at a given $a$, a particle scattered by a planet with a smaller $a_P$ will have a higher eccentricity than a particle scattered by a more distant planet.
Starting with a low-inclination population, the planets beyond $\sim 10 AU$ cannot efficiently produce particles with sufficiently high eccentricities to undergo Kozai oscillations at the semi-major axes of interest.
Therefore as long as Jupiter (and Saturn, to a lesser extent) were not far from their present day locations near the end of the gaseous disk's lifetime, the results are not sensitive to the other planets locations.
In these direct N-body simulation the potentially stabilizing impact of the 
Additionally it is important to note that the time $t=0$ in these calculations corresponds to the time of the formation of the giant planets, not the initial time of the formation of the disk.

\section{Comparison to Observations}\label{sec:observations}
In order to convert these retention fractions into a form suitable for comparisons with observations, we must make a series of assumptions about the properties of the initial population of planetesimals, specifically their initial abundance, size distribution, and physical characteristics.
Ideally we would like these assumptions to be guided by the observed characteristics of populations with the same source, unfortunately, we discuss later in this section, there is no observed population that we can be sure came from the same source region.
Therefore before we make a detailed comparison to observations we will first present an order of magnitude estimate for the number of high-inclination particles produced by this mechanism.

Theoretical models suggest that in order to form the cores of the giant planets within the observed lifetime of the gaseous protoplanetary disks, the initial disk mass must have been 2 to 5 or even 10 times larger than the minimum mass solar nebula (e.g. \citealt{Pollack.etal.1996,Lissauer.1987,Thommes.etal.2003,Alibert.etal.2005,Hubickyj.etal.2005} and see \citealt{Thommes.Duncan.2006} for a review).
This means that fewer than half of the initial planetesimals were incorporated into the giant planets, so they must have been ejected from the solar system or scattered into the sun.
Therefore, we assume that the total amount of mass in our solar system's initial planetesimal swarm must be \emph{at least} equal to the mass of solids incorporated into the planets, $\sim 40 M_\oplus$, and is probably significantly larger.

It is less clear how this mass was distributed among the planetesimals.
Without direct evidence for the true size distribution of these particles, 
we assume that the population has a size distribution similar to what is observed in other small body reservoirs in the outer solar system--an assumption we will revisit at the end of this section.
In general it appears that there is a relatively steep size distribution for large particles, which then, at some size $50 \lesssim D\lesssim 150$ km, rolls-over to a shallower distribution \citep{Bernstein.etal.2004,Fuentes.etal.2009,Fraser.Kavelaars.2009}.
Particles with sizes near this break dominate the mass of the whole population.
Therefore if we assume a similar size distribution for our initial population, this corresponds to greater than $10^{8}$ initial planetesimals of around $D= 100$ km in size.
With a capture efficiency around $10^{-3}$, we expect the mass in our hypothetical reservoir of objects with inclinations larger than 50$^\circ$ to be $>0.05 M_\oplus$, or $>10^5$ objects. 
This is of the same order as the upper limits on the mass of the observed Kuiper Belt \citep{Bernstein.etal.2004,Fuentes.Holman.2008} and to date there are no detected Kuiper belt objects with these inclinations.
Intuitively it seems unlikely that we would have failed to detect this massive of a population, implying that the primordial gas disk must have had a small radial extent to avoid producing this population in the first place.

\begin{figure}
\includegraphics[width=\columnwidth]{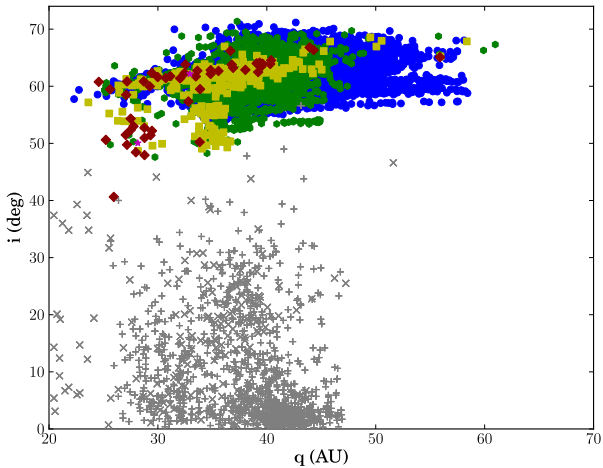}
\caption
{The perihelion distance v.~the inclination for all of the simulations (blue, green, yellow, and red symbols are the same as in figure \ref{fig:StabFig}) compared to the Kuiper Belt objects (+ symbols) and centaur/scattered disk objects (x symbols) from the minor planet center.}
\label{fig:q-i}
\end{figure}

To emphasize the differences between our hypothetical populations and the detected population we show the perihelion distance versus the inclination for all stable particles produced in all simulations as compared to the detected KBOs in figure \ref{fig:q-i}.
The observed Kuiper Belt occupies a distinctly different region of parameter space than any of the populations produced here.
We note that there are a few objects detected on the periphery of the area of parameter space that is populated by these models; for instance 2005 NU125 ($i=56.5^\circ$, $a=44.2$ AU, $e=0.026$).  
However, fewer than 1\% of particles produced in simulations with 100 AU disks have inclinations as low as $56.5^\circ$, and fewer than 5\% of particles produced by 80 AU disks have inclinations this low.
Therefore, if this object is a member of the population described in this paper we would expect to have detected 20-100 KBOs with higher inclinations.
Additionally the smaller 50 AU disks produce a few particles at lower inclination and smaller perihelia that could potentially be consistent with observed KBOs.
However these are still relatively low probability particles ($<10\%$) and, as we discuss later in this section, the current data sets place limits on these 50 AU size disks.

Using two surveys, the Deep Ecliptic Survey \citep[DES;][]{Millis.etal.2002} and the Palomar Survey \citep{Schwamb.etal.2010}, we can quantify this statement and make rigorous observational limits.
The DES survey observed over 800 deg$^2$ with a mean limiting VR magnitude 23.6 \citep{Elliot.etal.2005}.
Although this survey targeted the ecliptic ($\pm6^\circ$), it was still sensitive to these high inclination objects during a fraction of their orbits.
The Palomar survey is a shallower survey (mean limiting VR magnitude 21.3) but with a wider coverage of 12,000 deg$^2$ up to $\pm 40^\circ$ ecliptic latitude.
Using the $a$,$e$, and $i$ from our population of stable particles, we generate $N_{\rm test} = 10^5$ to $10^6$ test particles and create a synthetic population of orbits by assigning randomized orbital angles.
For each of these test particles we determine the probability ($p_{i,j}$) that each particle (indexed by $i$) will be detected on each survey field (indexed by $j$) as a function of its intrinsic brightness ($H$).
If a particle does not land on the field then $p_{i,j}(H)=0$.  
If it does land on the field, we can define a critical absolute magnitude ($H_{\rm crit}$), a function of both the orbital properties and detection limit on the observation,
\begin{equation}
	H_{\rm crit,i,j} \equiv m_{\rm limit,j} - 2.5 \log \pfrac{ d_i^2 r_i^2}{q(\chi_i)}.
\end{equation}
In this expression $d_i$ is the distance between the sun and object, and $r_i$ is the distance between the observer and object (both measured in AU)
and $m_{{\rm limit},j}$ is the background limiting magnitude for the field.
We parameterize the phase function ($q(\chi_i)$) of these small bodies with the H G formalism \citep{Bowell.etal.1989}, ie.
\begin{eqnarray}
	q_(\chi_i) &=& (1-G)\phi_1(\chi_i)+G\phi_2(\chi_i) \\ \nonumber
	\phi_1(\chi) &=& \exp\left(-3.33 \tan\pfrac{\chi}{2}^{0.63}\right) \\ \nonumber
	\phi_2(\chi) &=& \exp\left(-1.87 \tan\pfrac{\chi}{2}^{1.22}\right) 
\end{eqnarray}
where $\chi_i$ is the phase angle and $G$ is a parameter describing the strength of the opposition spike \citep[e.g.][]{Gehrels.1956,Hapke.etal.1993,Nelson.etal.1998,Rabinowitz.etal.2007}
However due to the fact that the fields are always chosen to view objects near opposition, even varying this parameter from $G$=0 to 1 changes the critical value by less than 0.2 magnitudes so it does not notably affect our conclusions.
While the probability of detection is a smoothed step function around this critical magnitude ($H_{\rm crit}$), due to the uncertainties in the population size distribution, a simple step function will be adequate for these estimates.
Therefore we use $p_{i,j}(H)  = 1 $ if $H < H_{\rm crit,i,j}$, and zero otherwise.
Additionally, in order to determine the orbital elements of the detected object it must be detected in at least two different observations.  
Therefore we define a new variable $\delta(j,j^\prime)$ which we set to 1 if the two observations are temporally separated such that that the observed motion of the object is large enough to be resolvable by the instrument and small enough that the observations are clearly of the same object, and  $\delta(j,j^\prime) =0$ otherwise.  
The total number objects we would expect to detect is then
\begin{eqnarray}
N_{\rm detect} &=& \frac{1}{N_{\rm test}} \displaystyle\sum_i^{N_{\rm test}}
\displaystyle\sum_j^{N_{\rm fields}}\displaystyle\sum_{j^\prime}^{N_{\rm fields}}
 \displaystyle\int_{H_{\rm min}}^{H_{\rm max}} p_{i,j}(H)p_{i,j^\prime}(H) \nonumber\\
 &&\times \delta(j,j^\prime) (1-f_{\rm obsc,j}) n(H) dH 
\end{eqnarray}
where $f_{{\rm obsc},i}$ is the small fraction of the field lost due to stellar obscuration, $n(H)\equiv dN/dH$ is the luminosity function for these hypothetical objects, and $H_{\rm min}$ and $H_{\rm max}$ are the intrinsic brightness range for these particles.

To continue we must make a number of assumptions about the properties of these hypothetical high-inclination objects.
For this analysis we assume that the initial planetesimals have a Kuiper belt-like size distribution.
\citet{Bernstein.etal.2004} found that the luminosity function can be well described by a combination of power-laws
\begin{equation}
n(H) = n_0 (10^{-\beta H}+ c 10^{-\beta^\prime H})^{-1},
\end{equation}
where $\beta$ is the slope on the bright end, $\beta^\prime$ is the slope on the faint end, and $c$ is chosen so that the break between these two populations matches observations.
Observations indicate at least two distinctly different populations in the outer solar system.
The dynamically hot Kuiper belt objects (KBOs) have a fairly shallow slope to their bright-end distribution, while the dynamically cold KBOs and the Jupiter Trojans have a steeper size distribution.
The different origins of these size distributions are not well understood.
There is evidence that the hot population may have accumulated closer to the sun \citep{Levison.Stern.2001,Gomes.2003,Levison.etal.2008}, where the protoplanetary disk had shorter accretion times.
For this reason we strongly prefer the assumption that the luminosity distribution of our hypothetical population is similar to that of the hot population, but we consider both for completeness.
We revisit this line of reasoning at the end of this section.

When looking at the bright end slope, 
\citet{Fraser.etal.2010a} found different values for this power-law index $\beta$ in different populations, $\beta_{\rm cold}=0.8$, and $\beta_{\rm hot}=0.35$.
However, the data cannot constrain the shallow end slope.  
\citet{Bernstein.etal.2004} found a break in both the hot and cold populations.
We include this break in the cold population and use $\beta_{\rm cold}^\prime=0.38$.
However, we follow \citet{Fraser.etal.2010a} and do not change the slope of the hot population after the ``break'' so $\beta_{\rm hot}^\prime=0.35$.

We assume that particles extend from 30 km to 1000 km in size with a break at $D=100$ km.
Additionally, we assume that the particle size distribution extends to 1 km for the estimates of the mass, but we note that the calculation is insensitive to this choice as most of the mass lies in larger particles.
To convert the number of expected detection into an estimate of the mass in this population we assume a bond albedo $p=0.05$ and a density of $\rho=2~{\rm g cm^{-3}}$.

Finally, we can create a synthetic population and use the previously described method to calculate the expected number of detections.
For example, in our fiducial disk model, $R_D=100$ AU, $\gamma=-1.5$, and $\Theta=\Theta_{\rm cut}$ the fractional trapping efficiency is greater than $2\times 10^{-3}$.
If the scattered particles have the same size distribution of the hot population, we expect the two surveys to have been able to detect $\sim 10^{-6}$ of the scattered objects.
Now, if we assume that the initial population of planetesimals was 40 $M_\oplus$, then, using the same size distribution, there would have been initially $10^{8}$ planetesimals greater that $D \gtrsim 30$ km.
Therefore if our solar system had had a 100 AU disk gaseous disk we would expect the these two surveys to have detected over 100 objects.
However, this number depends upon the initial assumption that we start out with a population of 40 $M_\oplus$ planetesimals, a conservative lower limit.

\begin{figure}
\includegraphics[width=\columnwidth]{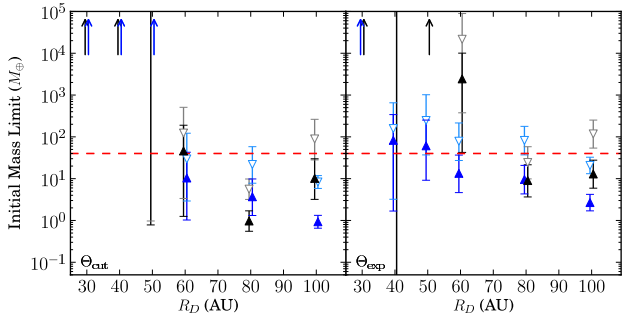}
\caption{
Observational constraints on the upper limit on the initial mass of planetesimals in the planet forming region.
As in figure \ref{fig:frequency_shape}, the left panel is for sharp cutoff disks and the right panel is for disks with exponential profiles.
The color indicates the disk surface density profile (On the left, $\gamma=1,3/2$ in blue and black, respectively. On the right, $\gamma=1/2,1$ in blue and black, respectively.)  
Upwards triangles indicate results for a shallow size distribution consistent with the hot population, while the downwards triangles indicate the steeper profile consistent with the cold population (see text).
Arrows indicate disk parameters for which we cannot limit the initial mass.
}
\label{fig:init_mass_both}
\end{figure}

So perhaps a better way to think about this is to 
turn the argument around and ask, for the different disk sizes, what is the upper limit on this initial mass.
Using a binomial distribution and a non-informative, Bayes-Laplace uniform prior we can determine the limit such that we are 95\% confident that we will detect at least one particle.
For the populations used here, that corresponds to $N_{\rm detect} \gtrsim $ 3.
The results are plotted in figure \ref{fig:init_mass_both}.
The upwards triangles indicate the results if we assume a size distribution consistent with the hot population (our preferred model), while the open downwards triangles indicate the limits assuming the size distribution of the cold population.
The error bars here are the statistical error bars (propagated through from figure \ref{fig:frequency_shape}).
For disks that do not produce stable high-inclination particles we can not place any limits, and these disks are indicated with upwards arrows.

If we assume that the size distribution of scattered planetesimals was similar to that of the hot population then we can rule out 80 and 100 AU disks as the limit on the initial population mass is below the expected theoretical amount.
Indeed, for disks with shallower profiles we can marginally rule out 60 AU disks with this size distribution.  
However, if we relax this assumption on the size distribution then we cannot place as stringent limits on the disks with steeper surface density profiles.

In order to significantly improve these constraints we would need a survey with a limiting magnitude similar to that of the DES (r=23.5) but with an area comparable to the Palomar survey (10,000 deg$^2$).
A non-detection in this type of survey would allow us to rule out all disks larger than 60 AU, except the steep exponential profile.
The LSST, designed to cover 10,000 deg$^2$ with a typical limiting magnitude of r=24.5 \citep{Ivezic.etal.2008} will easily exceed this limit, allowing us to make meaningful constraints down to the 50 AU level.

\subsection{Discussion of Size Distribution}
In this paper we argue that the non-detection of this hypothetical high-inclination population can be used to constrain the size of the protosolar nebula.
Of course, another possibility is that the population does exist, but that it has such a steep size-distribution that most of the mass is in bodies well below 100 km in size, and thus undetectable.
There are three main reasons why we believe this is unlikely.

First, there is another potential reservoir in which some remnants from this same initial population of planetesimals are believe to have been stored: the Sedna population.
The 1,000 km Sedna is so far unique in that it has a perihelion distance of 76 AU, well detached from Neptune, but it has a semimajor axis around 500 AU, well interior to the main Oort cloud \citep{Brown.etal.2004}.
The observational difficulties in detecting an object on this type of orbit implies there must be on the order of 100 similar sized planetoids on similar orbits.
The best dynamical mechanism to produce this type of orbit is to have an object scattered by a planet very early on in the solar system (at the same time considered in this paper) so that its pericenter can be lifted via an interaction with the birth cluster potential or by the passage of nearby stars \citep{Morbidelli.Levison.2004, Brasser.etal.2006}.
The large predicted number of 1,000 km-sized objects implies a shallow size distribution, and constraints from the TAOS survey indicate that this population must have $\beta < 0.88$\citep{Wang.etal.2009}, i.e. the size distribution cannot be much steeper than that of the cold population.

Second, in the Kuiper belt where we can directly measure the size distributions, we see that the hot population has a shallower size distribution than the cold population.
As mentioned previously, there is evidence that the hot population may have accumulated closer to the sun \citep{Levison.Stern.2001,Gomes.2003,Levison.etal.2008}, where the protoplanetary disk had shorter accretion times.
Therefore we would expect that the source region for our hypothetical population, being even closer to the sun than the hot population, may have a shallower population.
At least it seems unlikely to be significantly steeper.

Finally, we can be confident that the population that produced the comets could not have had a very steep size distribution.
If it had, then mutual collisions would have ground the comets down before they could have been scattered into the Oort Cloud \citep{Stern.Weissman.2001}.
\citet{Charnoz.Morbidelli.2003} demonstrated that this collisional grinding problem can be avoided by assuming a shallower size distribution up through 10s to 100s of km, such as the size distributions used in this paper. 
While the Oort cloud likely has a source region exterior to the population considered here \citep[the \emph{Nice} disk,][]{Tsiganis.etal.2005} or even from other stars \citep{Levison.etal.2010}, the same argument holds for the Sedna region.
For these reasons we believe that the size distribution of this population are unlikely to have a steeper distribution than the cold population and very likely have been similar to, or even more shallow than, the hot population.

\section{Conclusions and Discussion}\label{sec:conclusions}
In this paper we present a dynamical mechanism by which high inclination Kuiper Belt objects can be produced during the planet formation process due to Kozai cycles excited by the presence of the gaseous protoplanetary disk.
If our own solar system's disk had been radially extended ($R_D \gtrsim 50$ AU) a significant number of scattered planetesimals would have been placed onto high inclination orbits which would be stable for the lifetime of the solar system.
We find that the current observed lack of objects on these orbits in the DES and Palomar Distance Solar System Survey suggests that we can exclude very extended disks ($\gtrsim 80$ AU) so long as we are correct in our assumption of a size distribution more similar to the hot KBOs than the cold KBOs.
We argue that while we cannot technically rule out other options such as that the planetesimals in the planet forming region were orders of magnitude less numerous than expected by theory or that the size distribution was dominated by small planetesimals, we believe these options to be quite unlikely.

Future observations from deep all sky surveys (such as the LSST) will allow us to place more stringent constraints on the size of the disk.
With this type of survey, if we are able to detect a population of Kuiper belt objects with $i\sim60^\circ$ the orbital distribution of these particles will reveal information on the shape of the protoplanetary disks, especially revealing information about the truncation.
However, it is also possible (and perhaps likely) that future observations will not reveal this population.
If this is the case we will be able to conclude that the disk must have been smaller than 50 AU, but we will not be able to place more stringent limits because if the disk had been smaller than 50 AU, the dynamical mechanism presented in this paper cannot produce stable particles.

This method if the first technique to constrain the size of the gaseous protoplanetary disk of a known planetary system during the time of planet formation.
The disk sizes we find consistent with observations implies that in some ways our solar system is typical. 
Although many of the first protoplanetary disks observed were quite radially extended, smaller disks with sizes below 60 AU are quite common according to recent observations \citep{Andrews.etal.2010}.
Additionally, this small disk size is consistent with other pictures of the early solar system.
This work is consistent with the idea that a sharp cutoff in the planetesimal disk at around 50 AU might have corresponded to a sharp cut off in the gas disk.
Future deep all-sky surveys will allow us to probe closer to this 50 AU radius and allow us to meaningfully constrain on the theoretical models of of solid-gas interactions during the time of the formation of the solar system.

\acknowledgments
We thank M. Schwamb for guidance in our comparison with the Palomar Distant Solar System Survey.
KAK and HFL are grateful for funding from NASA's Origins program.

\end{document}